\documentstyle[aps,epsf]{revtex}

\draft

\begin{document}

\newcommand\BETA{{\rm Beta}}
\newcommand\<{\left<}
\renewcommand\>{\right>}


\title{Disoriented Chiral Condensates, Pion Probability Distributions
and Parallels with Disordered System}

\author{A.Z. Mekjian}
\address{Department of Physics, Rutgers University,
Piscataway, N.J.  08854}

\maketitle

\begin{abstract}
A general expression is discussed for pion probability distributions
coming from relativistic heavy ion collisions.
The general expression contains as limits:
1) The disoriented chiral condensate (DCC),
2) the negative binomial distribution and Pearson type III distribution,
3) a binomial or Gaussian result,
4) and a Poisson distribution.
This general expression approximates other distributions
such as a signal to noise laser distribution.
Similarities and differences of the DCC distribution with these
other distribution are studied.
A connection with the theory of disordered systems will be
discussed which include spin-glasses, randomly broken
objects, random and chaotic maps.
\end{abstract}




The purpose of this paper is to discuss a general expression for the
pion probability distribution which may be used to analyze pions coming
from relativistic heavy ion collisions.
The study of pions is of current interest for several reasons.
First pions are the main component of the produced particles
coming from such collisions.
Several thousand pions are now observed at CERN SPS experiments
and this number may go up by a factor of 10 at RHIC energies.
Secondly, the behavior of pions may signal the formation of
the quark-gluon plasma
as, for example, in the disoriented chiral condensate (DCC) picture
\cite{ref1,ref2,ref3,ref4,ref5,ref5a}.
Thirdly, the fluctuations in pions have been discussed in terms of
intermittency and fractal structure \cite{ref6} coming from non-Poissonian
effects.
A distribution which has been used to discuss these phenomena
is the negative binomial (NB) distribution \cite{ref7,ref8}.
The NB distribution also has an important feature known as
Koba-Nielsen-Olessen (KNO) scaling \cite{ref10} which the Poisson
distribution lacks.
Fourthly, the Bose-Einstein correlations amongst pions is an
important property used in Hanbury-Brown-Twiss (HBT) experiments,
and such correlations have also been proposed for the formation
of a pion laser \cite{ref11,ref12}.
Bose-Einstein condensation of atoms in a laser trap is a very
recently observed phenomena in another area of physics.
Distributions which extrapolate between Poisson and Bose-Einstein and
negative binomial have been developed,
such as the signal to noise model of Glauber-Lach.
This S/N model was originally developed in quantum optics,
but has also used for particle production and a review can be found
in \cite{ref9}.
The expression to be developed extends the range of extrapolation
by including not only these distribution but others.
Specifically, a general expression will be developed which contains
as special limiting cases the DCC distribution, the NB distribution,
a binomial distribution \cite{ref13} or Gaussian like distribution,
and the Poisson distribution.
The general distribution also approximates the S/N model.
A connection with the theory of disordered systems \cite{ref14,ref15}
will also be discussed further.
These disordered systems included spin-glasses \cite{ref16,ref17},
randomly broken objects \cite{ref18,ref19}, random permutations \cite{ref20},
random maps \cite{ref15,ref18}, and chaotic maps \cite{ref21}.
For example, the disoriented chiral condensate gives a probability
distribution which comes from a random direction of the isospin vector
and a connection of the DCC state with the theory of disordered systems
will be noted.
The probability distribution is simply given by
\begin{eqnarray}
 P_k(N, x, \gamma) = \pmatrix{N \cr k}
          \frac{\BETA[N-k+x, k+\gamma]}{\BETA[x,\gamma]}
                \label{eq1}
\end{eqnarray}
where $k$ is a random variable which can take on values
$k = 0$, 1, 2, ..., $N$, and $x$, $\gamma$ are parameters.
The $\BETA$ function that appears in Eq.(\ref{eq1}) is
$\BETA[\omega, z] = \Gamma(\omega)\Gamma(z)/\Gamma(\omega+z)$
where $\Gamma(z)$ is a gamma function.
In probability theory, $P_k(N, x, \gamma)$ is known as a Poly\'{a}
distribution \cite{ref23}.
This distribution was used in Ref.\cite{ref24} to describe
cluster yields coming from high energy heavy ion collisions
and a connection of it with the theory of disordered systems was
discussed briefly in Ref.\cite{ref14}.
Its usefulness as a model for pion probability distribution will
be developed.

Some interesting limits of Eq.(\ref{eq1}) are as follows.
Setting $x=1$, $\gamma=1/2$ in Eq.(\ref{eq1}),
the $P_k(N, x, \gamma) = P_k(N, 1, 1/2)$ is given by
\begin{eqnarray}
 P_k(N, 1, 1/2) = \frac{(N!)^2 2^{2N} (2k)!}{(2N+1)! (k! 2^k)^2}
    \sim \frac{1}{2\sqrt{N k}} .         \label{eq2}
\end{eqnarray}
Eq.(\ref{eq2}) appears in pionic yields from the disoriented chiral
condensate model of a QCD phase transition.
Specifically $2k = n_0$ is the number of neutral pions
and $2N = n_0 + n_+ + n_-$ is the total number of pions with
$n_+ = n_-$ being the number of positive or negative pions.
In obtaining $1/2\sqrt{Nk}$, Stirling's approximation was used.
The probability distribution for the neutral to total pion yield, 
$R_3 = n_0/(n_0 + n_+ + n_-)$, is then $P(R_3) \sim 1/2\sqrt{R_3}$.
A general expression for the fluctuation of Eq.(\ref{eq1}) is \cite{ref14}
\begin{eqnarray}
 \<k^2\> - \<k\>^2 = \frac{x}{x+\gamma+1} \<k\>
       \left(1 + \frac{\<k\>}{\gamma}\right)
         \label{eq3}
\end{eqnarray}
where $\<k\> = \left(\gamma/(x+\gamma)\right) N$.
For $x=1$, $\gamma=1/2$, $\<k\> = (1/3) N$ or
 $\<n_0\> = 2N/3 = \<n_+\> = \<n_-\>$.
Thus $\<n_0^2\> - \<n_0\>^2 = (4/5) \<n_0\> (1 + \<n_0\>)$ and
$\<n_+^2\> - \<n_+\>^2 =
  (1/5) \<n_+\> (1 + \<n_+\>) = (1/4) (\<n_0^2\> - \<n_0\>^2)$.
The probability distribution of $j$ $\pi_+$'s or $\pi_-$'s is
\begin{eqnarray}
 P_j(N, 1, 1/2) = \frac{(N!)^2 2^{2j}}{(2N+1)!}
            \frac{(2(N-j))!}{((N-j)!)^2} 
     \sim \frac{1}{2\sqrt{N(N-j)}}         \label{eq4}
\end{eqnarray}
The $P_j(N, 1, 1/2)$ increases with $j$ and behaves asymptotically
like the arcsine distribution $1/\pi\sqrt{j(N-j)}$ for $j \sim N$
when $N$ is very large.
This behavior is totally different than the $\pi_0$ behavior
simply obtained from Eq.(\ref{eq2}).
As will be shown the $\pi_0$ behavior can be approximated by other
distributions, but not this $\pi_+$ or $\pi_-$ behavior which is
coupled to the $\pi_0$ distribution.
The arcsine distribution is also a special case of a distribution related
to Eq.(\ref{eq1}) to be discussed below.
The constraint $2N = n_0 + n_+ + n_-$ represents a major difference between
the DCC model and a NB description and other probability distributions
usually applied to pionic yields.

A binomial limit of Eq.(\ref{eq1}) is obtained in the limit
$x \to \infty$, $\gamma \to \infty$.
If $x = 2\gamma$ so that $\<n_0\> = \<n_+\> = \<n_-\> = \frac{1}{3} (2N)$,
then
 $P_k(N) = \pmatrix{N \cr k} \left(\frac{1}{3}\right)^k
    \left(\frac{2}{3}\right)^{N-k}$.
Also $\<k^2\> - \<k\>^2 = (2/3) \<k\>$ from Eq.(\ref{eq1}) and
since $n_0 = 2k$: $\<n_0^2\> - \<n_0\>^2 = (4/3) \<n_0\>$.
For large $N$, the binomial distribution is approximately a Gaussian.
The DCC distribution for a large number of domains is also a binomial.
In the limit in which each domain has $k = 0$ or 1 or $n_0 = 0$, 2 $\pi_0$'s
so that the number of domains equals $N$, then the $P_k(N)$ is also
given by the binomial result quoted above.
Moreover, the DCC distribution has the following central limit theorem feature.
Let $m$ equal the number of domains and let $N_1$ be the maximum $k_1$
in any one domain, i.e., $k_1 = 0$, 1, 2, ..., $N_1$,
with $n_{01} = 0$, 2, ..., $2N_1$ the random number of $\pi_0$'s
coming from each domain.
The fluctuation in one domain is $\<k_1^2\> - \<k_1\>^2 = \sigma_1^2$
and is given by Eq.(\ref{eq3}) with $x=1$, $\gamma=1/2$,
or using $\<n_1\> = N_1/3$: $\<k_1^2\> - \<k_1\>^2 = (2/15) N_1 (1 + 2N_1/3)$.
For $m$ DCC cells, the total variance is
 $\sigma_m^2 = m \sigma_1^2 = (2/15) N (1 + 2N/3m)$.
When $m = N$, $\sigma_m^2 = (2/9)N$ which is the binomial result.

A negative binomial limit can also be obtained from Eq.(\ref{eq1})
in the limit $N \to \infty$ $x \to \infty$, $\rho = x/N$
and $p = \rho/(1+\rho)$. Then
\begin{eqnarray}
 P_k(p, \gamma) &=& \pmatrix{\gamma+k-1 \cr k} p^\gamma (1-p)^k  \nonumber \\
  &=& \pmatrix{\gamma+k-1 \cr k} \frac{1}{\left(1 + \<k\>/\gamma\right)^\gamma}
      \left(\frac{\<k\>/\gamma}{1 + \<k\>/\gamma}\right)^k        \label{eq5}
\end{eqnarray}
where $\<k\> = \left(\gamma/(x+\gamma)\right) N = \gamma/\rho$
and $\<k^2\> - \<k\>^2 = \<k\> (1 + \<k\>/\gamma)$ from Eq.(\ref{eq3}).
The $P_k$ of Eq.(\ref{eq5}) is the NB distribution with $\gamma$ now
the NB parameter.
The NB distribution is of current interest since this distribution fits
many experimental observations as, for example, in the multiplicity
distribution of hadrons produced in $e^+ e^-$ collisions at LEP
and PEP-PETRA and in hadron and heavy ion collisions \cite{ref25}.
Van Hove and Giovannini \cite{ref7,ref8} have discussed
a clan model which gives a NB distribution.
Carruthers and Shih \cite{ref9} summarized various mechanisms that lead to
NB distributions and show that this distribution can be put into a
correspondence with self-similar Cantor sets or fractal structures.
The NB distribution has been used to discuss issues related to
intermittency \cite{ref6,ref8}.
Taking $\gamma = 1/2$, the same value of $\gamma$ as in the DCC
distribution, then
\begin{eqnarray}
 P_k(1/2) = \frac{(2k)!}{2^k (k!)^2} \left(\frac{1}{1 + 2\<k\>}\right)^{1/2}
            \left(\frac{2\<k\>}{1+2\<k\>}\right)^k        \label{eq6}
\end{eqnarray}
The factor $(2k)!/2^k(k!)^2$ also appears in the DCC distribution and
its Stirling limit is $1/\sqrt{\pi k}$ which gives the $1/\sqrt{R_3}$
behavior associated with the DCC state.
The factor $\left(2\<k\>/(1 + 2\<k\>)\right)^k$ gives a geometric series
decrease or exponential decrease in $P_k$.
Writing $\left(2\<k\>/(1 + 2\<k\>)\right) = e^{-\alpha k}$ with
$\alpha = \ln(1 + 1/2 \<k\>) \approx 1/2 \<k\>$ results in
$P_k(1/2) \approx \left(1/\sqrt{2\pi k \<k\>}\right) e^{-k/2\<k\>}$.
For general $\gamma$,
\begin{eqnarray}
 P_k(\gamma) \approx \left(k^{\gamma-1}/\Gamma(\gamma)\right)
      \left(\gamma/\<k\>\right)^\gamma e^{-\gamma k/\<k\>}
                \label{eq7}
\end{eqnarray}
which is a Peason type III distribution.

The Poisson limit of the NB is realized when $\gamma \gg \<k\>$.
Both the NB and DCC have larger than Poisson fluctuations.
Fluctuations of pions larger than Poisson are to be expected just on
the basis of Bose-Einstein statistics, with the Poisson limit
obtained from Maxwell-Boltzmann statistics.
A 20\% enhancement in fluctuations above those of Poisson statistics
occurs in both hydrodynamic and thermodynamic models from
Bose-Einstein correlations \cite{ref26}.
An important feature of the NB and Peason type distribution is
that $\<k\> P_k(\gamma)$ is a universal function of the scaled
variable $k/\<k\>$ in the limit $k \to \infty$, $\<k\> \to \infty$,
which is known as KNO scaling \cite{ref10}.
For example the Peason Type III is simply
 $k P_k(\gamma) = \left(\gamma^\gamma / \Gamma(\gamma)\right)
    \left(k/\<k\>\right)^{\gamma-1} e^{-\gamma k / \<k\>}$
and the DCC distribution is
 $\<k\> P_k \cong \left(1/2\sqrt{3}\right) \left(\frac{\<k\>}{k}\right)^{1/2}$.
However, the Poisson distribution does not obey KNO scaling.

Other distribution have been developed which extrapolate between
Poisson and NB and have been used to describe the multiplicity
distribution of produced particles.
These include the Glauber-Lach signal S to noise NL model \cite{ref9} with S
the coherent signal level (Poisson emitter) and NL the thermal
Bose-Einstein noise level.
In Biyajima's generalization of this model \cite{ref9,ref27},
the probability of $k$
particles is given by a distribution which is called a laser distribution:
\begin{eqnarray}
 P(k) = \frac{(NL/\alpha)^k}{(1+NL/\alpha)^{k+\alpha}}
     e^{- S/\left(1+\frac{NL}{\alpha}\right)}
   L_k^{\alpha-1} \left(\frac{-\alpha\frac{S}{NL}}{1+\frac{NL}{\alpha}}\right)
         \label{eq8}
\end{eqnarray}
with $\alpha=1$ the Glauber-Lach model and $L_k^{\alpha-1}$ is a
generalized Laguerre function.
The $\<k\> = NL + S$ and $\<k^2\> - \<k\>^2 = \<k\> + NL (2 S + NL)/\alpha$.
The Poisson limit is achieved when $NL \to 0$,
while a NB limit follows for $S \to 0$.
By comparison, the distribution of Eq.(\ref{eq1}) also contains
these two limits as well as the DCC distribution and other distributions
already mentioned.
Fig.1 illustrates the similarity of Eq.(\ref{eq1}) for the DCC choice
of $x = 1$, $\gamma = 1/2$ and for $N = 300$ with the distribution
of Eq.(\ref{eq8}) with $\alpha = 1/2$, $S = 62$, $NL = 38$.
Both distributions have $\<k\> = 100$.
In the DCC choice $k$ is restricted to $0 \le k \le N$
while for signal to noise models $0 \le k \le \infty$.
The figure shows the interval $0 \le k \le 300$ only.
As noted, a major difference between the DCC results and these other models
is the constraint $2N = \pi_0 + \pi_+ + \pi_-$.
This leads to totally different behaviors in the $\pi_0$ channel and
the $\pi_+$, $\pi_-$ channel.
This result suggests that event-by-event data should be investigated
not only for the fall off in $\pi_0$ to total yield as $1/2\sqrt{R_3}$
but also the rise in the $\pi_+$ or $\pi_-$ yields as given by Eq.(\ref{eq4}).
Also shown in this figure is a NB distribution
with $\gamma = 1/2$ and $\<k\> = 100$.
If the fluctuations of the two distributions, Eq.(\ref{eq1})
and Eq.(\ref{eq8}) are equated then $\alpha$, $\gamma$ are connected
by $\gamma = \alpha/(1 - b^2)$ where $b \equiv S/(NL + S)$ is the
fractional signal level.
Very large Bose-Einstein enhancements to Poisson results can appear
in other models such as the pion laser model of Ref.\cite{ref11}.

\begin{figure}[tbp] \label{fig1}
\centerline{
  \epsfxsize=3.0in
  \epsfbox{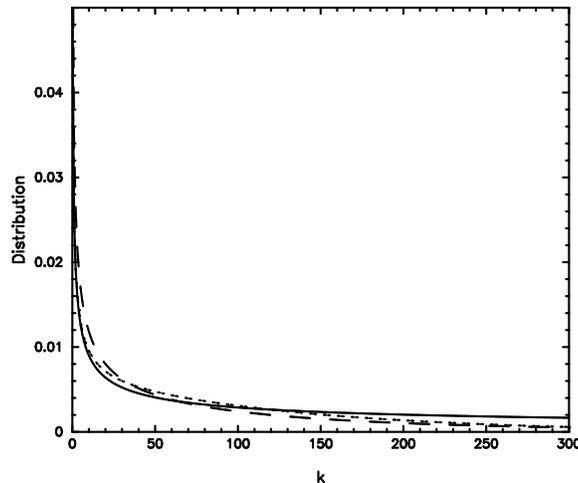}
 }
\caption{
The distribution of Eq.(\protect\ref{eq1}) for the DCC choice
$x = 1$, $\gamma = 1/2$ (solid line) compared with the negative binomial
distribution with $\gamma = 1/2$ and $\<k\> = 100$ (dashed line)
and a signal to noise model with $S = 62$, $NL = 38$ and $\alpha = 1/2$
(dotted line). The three distributions have similar features.}
\end{figure}

When Eq.(\ref{eq1}) is rewritten as
\begin{eqnarray}
 P_k(N, x, \gamma) = \int \pmatrix{N \cr k} w^k (1 - w)^{N-k}
      u(w, x, \gamma) dw   \label{eq9}
\end{eqnarray}
important functions as
 $u(w,x,\gamma) = \left(\BETA(x,\gamma)\right)^{-1} w^{\gamma-1} (1-w)^{x-1}$
and $f(w, x, \gamma) = w^{-1} u(w)$ emerge.
The $u(w,x,\gamma)$ is also the limit of $P_k(N,x,\gamma)$
as $N \to \infty$, $k \to \infty$, $k/N \to w$.
The Stirling limit of Eq.(\ref{eq1}) corresponds to $u(w,1,1/2)$.
The distribution of Eq.(\ref{eq9}) and $u(w,x,\gamma)$
and its associated $f(w,x,\gamma)$ play a prominant role in the theory
of disordered systems such as spin-glasses, randomly broken objects,
random and chaotic maps, etc.
Table 1 summarise some of these connections.
The choice of $x$ and $\gamma$ varies for the different areas as listed in
the table. The last column gives the quantity described by Eq.(\ref{eq9}).
In spin-glass models, random hamiltonians based on an Ising interaction
$J \vec\sigma_i\cdot\vec\sigma_j$ (where $J = \pm |J|$ is chosen randomly)
are used to calculate rugged free-energy landscapes.
The function $f(w,x,\gamma)$ and its related probability function
$u(w,x,\gamma)$ give the distribution of well depths in the
free energy landscape.
For clusters and breaking processes, with initial size $A$ into fragments
of size $k = 1$, 2, ..., $A$ with $n_k =$ the number of fragments of size $k$,
$f(w,x,\gamma)$ is related to $n_k$ by $A n_k \to f(w,x,\gamma)$ in the
limit $k$, $A \to \infty$ with $w = k/A$.
At $x = 1$, $n_k \sim 1/k^\tau$, a power law distribution with
Fisher exponent $\tau = 2 - \gamma$.
In the case of random permutations in Table 1, the $n_k$ of the cluster
example just given now becomes the number of cycles of length $k$,
when the symmetric group of permutations is represented by its cycle
class structure.
A cycle class description for pionic distributions was also described
in Ref.\cite{ref27}.
The random map example of Table 1 is also a prototype model
for disordered systems \cite{ref19}.
Here $n_k$ is the number of attractors with $k$ elements.
These attractors are limit cycles produced in a random mapping of
$A$ points into itself (allowing multiple points to go into one point).
The $f(w,x,\gamma)$ and $u(w,x,\gamma) = w f(w,x,\gamma)$
follow as in the cluster case above.
Chaotic maps, such as the tent and quadratic map \cite{ref22,ref23},
have invariant distributions which are given by arcsine laws.
The $u(w,x,\gamma)$ for $x = \gamma = 1/2$ is the
arcsine distribution $u(w,x,\gamma) = \pi/\left(w(1-w)\right)^{1/2}$.
%

%
%
\begin{table} \label{tabl1}
\caption{Various disordered systems and values for $x$, $\gamma$
described by Eq.(\protect\ref{eq1}).}
\begin{center}
\begin{tabular}{|l|cc|l|}
 \ \ \ \ \ \ \ \  Area  &  $x$  &  $\gamma$   & \ \ \ \ \ Quantity Described \\
\hline 
            &    &    &     \\
 Spin Glass \cite{ref16,ref17}
    & \multicolumn{2}{c|}{\ \ $x + \gamma = 1$ \ \ }
    & Well depths from random hamiltonians \\
            &    &    &     \\
 Disoriented Chiral Condensate - DCC \cite{ref1,ref2,ref3,ref4,ref5}
    &  1   &  1/2  
    & Pion probability distribution  \\
            &    &    &     \\
 Randomly Broken Objects \cite{ref15,ref18,ref19}
    & $x$  &   1   
    & Fragment sizes  \\
            &    &    &     \\
 Random Permutations \cite{ref20}
    &  1   &   1  
    & Cycle length      \\
            &    &    &     \\
 Random Maps \cite{ref15,ref18}
    & 1/2  &   1 
    & Limit cycle distribution        \\
            &    &    &     \\
 Chaotic Maps; Tent and Quadratic \cite{ref21,ref22}
    & 1/2  &  1/2  
    & Invariant distribution    \\
            &    &    &     \\
 Power Laws with Fisher Exponent \cite{ref14}
    &  1   & $\gamma$ 
    & Cluster or droplet distributions      \\
 \ \  $\tau < 2$ and $\tau = 2 - \gamma$  &  &  &  \\
            &    &    &     \\
%
\end{tabular}
\end{center}
\end{table}
%

%

To summarize and conclude, this paper contained a general expression
for pions coming from relativistic heavy ion collisions.
Various limits of this general distribution contain frequently used
pion distribution.
One important example is the disordered chiral condensate distribution
which is of current interest because of its use as a possible signal
of the quark-gluon phase transition.
Another important limit gives rise to the negative binomial
distribution which appears in discussions of intermittency,
KNO scaling, non-Poissonian fluctuations.
Other limits include a binomial limit, or Gaussian like limit,
and a Poisson limit.
Distributions which extrapolate between Poisson and negative binomial such
as the Glauber-Lach and laser distribution have been developed
previously and this paper extends the range of limiting cases to include
many other distributions.
Comparisons are made between the DCC, NB distribution, and laser
distribution, with each having large fluctuations.
Similarities and differences which may be important in understanding the
quark-gluon phase transition are discussed.
For example, the DCC distribution has a probability distribution
of $\pi_0$'s which falls a $1/\sqrt{k}$, Eq.(\ref{eq2}), coupled
with a $\pi_+$ or $\pi_-$ distribution which rises as $1/\sqrt{N-k}$,
Eq.(\ref{eq4}), as $j \to N$ for $j$ $\pi_+$'s or $\pi_-$'s.
The $\pi_0$ distribution can be approximated by a standard negative
binomial model or a signal to noise laser model as shown in Fig.1.
Thus, this unique coupling of the $\pi_0$ and $\pi_+$ or $\pi_-$ channel
characterizes the DCC state more than just the often quoted $1/\sqrt{R_3}$
or $1/\sqrt{k}$ behavior in the probability distribution of $\pi_0$'s.
Heavy ion pion production data may also be analyzed in terms of the general
distribution.
The connection of this general distribution with the theory of disordered 
systems is discussed.
These connections included spin glasses, randomly broken objects,
random permutations, random maps and chaotic maps.

This research was supported by a DOE grant, Grant \# DE-FG02-96ER-40987.

\end{document}